\documentclass[proceedings]{stacs}
\stacsheading{2010}{95-106}{Nancy, France}
\firstpageno{95}

\theoremstyle{plain}
\theoremstyle{definition}

\begin{document}

\title{Exact Covers via Determinants}

\author{A.~Bj\"orklund}{Andreas Bj\"orklund}
\address{}
\email{andreas.bjorklund@yahoo.se}
\date{}


\begin{abstract}
Given a $k$-uniform hypergraph on $n$ vertices, partitioned in $k$ equal parts such that every hyperedge includes one vertex from each part,
the $k$-Dimensional Matching problem asks whether there is a disjoint collection of the hyperedges which covers all vertices.
We show it can be solved by a randomized polynomial space algorithm in $O^*(2^{n(k-2)/k})$ time. The $O^*()$ notation hides factors
polynomial in $n$ and $k$.

The general Exact Cover by $k$-Sets problem asks the same when the partition constraint is dropped and arbitrary hyperedges of cardinality $k$ are permitted.
We show it can be solved by a randomized polynomial space algorithm in $O^*(c_k^n)$ time, where $c_3=1.496, c_4=1.642, c_5=1.721$, and provide a
 general bound for larger $k$.

Both results substantially improve on the previous best algorithms for these problems, especially for small $k$. They follow from the new observation
that Lov\'asz' perfect matching detection via determinants (Lov\'asz, 1979) admits an embedding in the recently proposed inclusion--exclusion
counting scheme for set covers, \emph{despite} its inability to count the perfect matchings.
\end{abstract}
\maketitle 

\section{Introduction}
The Exact Cover by $k$-Sets problem (X$k$C) and its constrained variant $k$-Dimensional Matching ($k$DM) are two well-known NP-hard problems.
They ask, given a $k$-uniform hypergraph, if there is a subset of the hyperedges which cover the vertices without overlapping each other.
In the $k$DM problem the vertices are further partitioned in $k$ equal parts and the hyperedges each includes exactly one vertex from each part.
While being two of the 21 items of Karp's classic list of NP-complete problems~\cite{K72} for $k\geq 3$, little is known on their algorithmic side.
In this paper, we present stronger worst case time bounds for these problems by combining Lov\'asz' perfect matching detection algorithm
via determinants \cite{L79} with the inclusion--exclusion counting for set covers \cite{BH08}. We show

\begin{thm}
\label{thm: kDM}
$k$-Dimensional Matching on $n$ vertices can be solved by a Monte Carlo algorithm with exponentially low probability of failure in $n$, using space polynomial in $n$,
running in
$
O^*(2^{n(k-2)/k})
$
time.
\end{thm}
\begin{thm}
\label{thm: XkC}
Exact Cover by $k$-Sets on $n$ vertices can be solved by a Monte Carlo algorithm with exponentially low probability of failure in $n$, using space polynomial in $n$,
running in
$
O^*(c_k^n)
$
time, 
with $c_3=1.496, c_4=1.642, c_5=1.721, c_6=1.771, c_7=1.806$, 
and in general $c_k<2\left(8.415k^{0.9-k}(k-1)^{0.6}(k-1.5)^{k-1.5}\right)^{-1/k}$
\end{thm}

These bounds are large improvements over the previously known ones.
In particular, for three dimensional matching our algorithm runs in time asymptotically proportional to the square root of the previous best algorithm's runtime.

We hope the present paper conveys the message that inclusion--exclusion is amendable not only to counting problems, but can at times be used more directly to settle the decision version of a problem.

\subsection{Previous Work}
Perhaps the most famous algorithmic contribution on the subject of exact covers is Knuth's \emph{Dancing Links} paper \cite{K00}, which actually just addresses
a general implementation issue which saves a small constant factor in the natural backtracking algorithm for the problem. 
About the backtracking approach on exact cover he writes ``Indeed, I can't think of any other reasonable way to do the job in general''. 
While we certainly may agree depending on how
much you put in the words ``reasonable'' and ``general'', we must point out that the best provable worst case bounds for the problems are obtained by
analyzing very different algorithms. 
 Bj\"orklund et al. \cite{BHK09} uses inclusion--exclusion and fast zeta transforms on the full subset lattice to show that exact set covers 
 of any $n$ vertex hypergraph can be counted in $O^*(2^n)$ time even when the number of hyperedges to choose from are exponential.
Restricted to $k$-uniform hypergraphs, Koivisto \cite{K09} proposes a simple clever dynamic programming over subsets which show that Exact Cover by $k$-Sets
can be solved in $O^*(2^{n(2k-2)/\sqrt{(2k-1)^2-2\mbox{ln}(2)}})$ time. The algorithm is actually capable of counting the solutions 
and also works for not necessarily disjoint
covers. It does, however, use exponential space. The best previous algorithm for the problem using only polynomial space is given in \cite{BH08} and has a runtime bound in $O^*((1+k/(k-1))^{n(k-1)/k})$.
For $k$-Dimensional Matching, the best known algorithm as far as we know is an $O^*(2^{n(k-1)/k})$ time algorithm
resulting from a generalization of Ryser's inclusion--exclusion counting formula for the permanent \cite{R63}, presented in \cite{BH08}. 
A comparison of the bounds guaranteed by these algorithms and the ones given in this paper is shown in Table~\ref{tab: runtime} for small $k$.

\begin{center}
    \begin{table}
    \begin{tabular}{ | l | l | l | l | l | l | l |}
    \hline
     Algorithm  $\backslash \, k$     &     3 &     4 &     5 &     6 &     7 &     8 \\ \hline
    $k$DM in \cite{BH08} & 1.587 & 1.682 & 1.741 & 1.782 & 1.811 & 1.834 \\ \hline
    $k$DM here           & 1.260 & 1.414 & 1.516 & 1.587 & 1.641 & 1.682 \\ \hline
    X$k$C in \cite{BH08} & 1.842 & 1.888 & 1.913 & 1.929 & 1.940 & 1.948 \\ \hline
    X$k$C in \cite{K09}  & 1.769 & 1.827 & 1.862 & 1.885 & 1.901 & 1.914 \\ \hline
    X$k$C here           & 1.496 & 1.642 & 1.721 & 1.771 & 1.806 & 1.832 \\ \hline
   
    \end{tabular}
   
    \caption{ Comparison of the base $c$ in the $O^*(c^n)$ runtime of previous and the new algorithms.}
    \label{tab: runtime}
    \end{table}
\end{center}
 
For $k=2$ the problems X$2$C and $2$DM are better known as the problems of finding a perfect matching in a general and bipartite graph, respectively. For these problems several polynomial time algorithms are known.
We definitely admit that it seems like an obvious idea to try to reduce the $k>2$ cases to the $k=2$ case searching for faster algorithms for larger $k$. Still, we
believe that it is far from clear how to achieve this efficiently. In this paper we make such an attempt by reducing the $k>2$ cases to variants of one of the first polynomial time algorithms for detecting the existence of perfect matchings: 
Lov\'asz' algorithm from \cite{L79} which evaluates the determinant of the graph's Tutte matrix \cite{T47} at a random point.

\section{Our Approach}

\subsection{Preliminaries}
We use the terminology of (multi)hypergraphs. A hypergraph $H=(V,E)$ is a set $V$ of $n$ vertices and a \emph{multiset} $E$ of (hyper)edges 
which are subsets of $V$. Note in particular that with this definition edges may include only one (or even no) vertex and may appear more than once. In a $k$-uniform hypergraph each edge $e\in E$ has size $|e|=k$.
Given a vertex subset $U\subseteq V$, the \emph{projected hypergraph} of $H=(V,E)$ on $U$, denoted $H[U]=(U,E[U])$ is a hypergraph on
$U$ where there is one edge $e_U$ in $E[U]$ for every $e\in E$, defined by $e_U=e\cap U$, i.e. the projection of $e$ on $U$.

We study two related problems.

\begin{definition}[$k$-Dimensional Matching, $k$DM]
\
\begin{description}
\item[Input] A $k$-uniform hypergraph $H=(V_1\cup V_2\cup \cdots V_k,E)$, with $E\subseteq V_1\times V_2\times \cdots V_k$.
\item[Question] Is there $S\subseteq E$ s.t. $\cup_{s\in S} s = V_1\cup V_2\cup \cdots V_k$ and $\forall s_1\neq s_2\in S: s_1\cap s_2=\emptyset$.
\end{description}
\end{definition}

\begin{definition}[Exact Cover by $k$-Sets, X$k$C]
\
\begin{description}
\item[Input] A $k$-uniform hypergraph $H=(V,E)$.
\item[Question] Is there $S\subseteq E$ s.t. $\cup_{s\in S} s = V$ and $\forall s_1\neq s_2\in S: s_1\cap s_2=\emptyset$.
\end{description}
\end{definition}

For a matrix $\mathbf{A}$ we will by $\mathbf{A}_{i,j}$ denote the entry at row $i$ and column $j$.

\subsection{Determinants}
The determinant of an $n\times n$-matrix $\mathbf{A}$ over an arbitrary ring $R$ can be defined by the Leibniz formula
\begin{equation}
\label{eq: det}
\mbox{det}(\mathbf{A})=\sum_{\sigma:[n]\rightarrow [n]} \mbox{sgn}(\sigma)\prod_{i=1}^n \mathbf{A}_{i,\sigma(i)}
\end{equation}
where the summation is over all permutations of $n$ elements, and $\mbox{sgn}$ is a function called the sign of the permutation which assigns either one or minus one
to a permutation.
In this paper we will restrict ourselves to computing determinants over fields of characteristic two, GF($2^m$) for some positive integer $m$.
In such fields every element serves as its own additive inverse, and in particular so does the element one, and the $\mbox{sgn}$ function identically maps one to every permutation.
Thus it vanishes from Eq.~\ref{eq: det} in this case, and the determinant coincides with another matrix quantity, called the \emph{permanent}:
\begin{equation}
\label{eq: per}
\mbox{per}(\mathbf{A})=\sum_{\sigma:[n]\rightarrow [n]} \prod_{i=1}^n \mathbf{A}_{i,\sigma(i)}
\end{equation}
Permanents of $0$--$1$-matrices over the natural numbers are known to count the perfect matchings of the bipartite graph described by the matrix. The reader may subsequently be tempted to think that this identity of determinants and permanents over fields of characteristic  two is the property that makes our algorithms work. There is however nothing magical about these fields in this context. Our reason for working in GF($2^m$) is simply that with this choice of fields we don't even have to define the sign function, making several of the proof arguments later on much easier to digest. In principle though, \emph{any} large enough field will work, with slightly more complicated proofs.

The interesting property of the determinant that we will exploit here is that although it is defined above in Eq.~\ref{eq: det} as a sum of an exponential number of terms,
it admits computation in time polynomial in $n$. This can be achieved for instance via the so called LU-factorization of the matrix which almost any textbook on linear algebra will tell you. In fact, computing the determinant is no
harder than square matrix multiplication, see \cite{BH74}, and hence it can be done in $O(n^\omega)$ field operations where $\omega=2.376$ is the Coppersmith--Winograd
exponent \cite{CW}.

\subsection{Inclusion--Exclusion for Set Covers}
Let us review the inclusion--exclusion counting scheme for exact set covers presented by Bj\"orklund and Husfeldt in \cite{BH08}:
Given a $k$-uniform hypergraph $H=(V,E)$ and any subset $U\subseteq V$, we can count the number of Exact Covers by $k$-Sets, denoted $\#\mbox{X}k\mbox{C}(H)$, by the inclusion--exclusion 
formula
\begin{equation}
\label{eq: inc-exc}
\#\mbox{X}k\mbox{C}(H)=\sum_{X\subseteq V-U} (-1)^{|X|} W(H,U,X)
\end{equation}
where $W(H,U,X)$ counts the number of ways to exactly cover $U$ with $|V|/k$ edges in $H[U]$ whose corresponding edges in $H$ are disjoint from $X$.
Put differently, $W(H,U,X)$ counts the number of ways to pick $|V|/k$ edges from $H$, all having an empty intersection with $X$, which cover $U$ without any overlap.
In particular, when $U=\emptyset$ it is straightforward to compute $W(H,\emptyset,X)$ by just counting the number of edges in $H$ disjoint from $X$, calling this quantity $d(X)$,
and then computing the binomial $\binom{d(X)}{m}$. In \cite{BH08}, some examples where this algorithm could be accelerated by choosing a larger $U$ were identified
where the speedup was obtained by utilizing $U$'s such that the projected hypergraph on $U$ had low path--width. This enabled efficient counting by dynamic programming over a path decomposition.

\subsection{Moving to GF($2^m$)}
In this paper, we find a new way to allow a large $U$ to expedite the computation of the formula Eq.~\ref{eq: inc-exc} above. We observe that whenever the projected hypergraph
contains edges of size at most two, we can use determinants to compute the formula faster. 
We note that if the problem of counting perfect matching had an efficient algorithm $A$, we would almost immediately get an $O^*(2^{n(k-2)/k})$ time algorithm for the $k$DM problem. We would simply let $U$ be any two of the parts in the input partition, and use $A$ to compute $W(H,U,X)$. Unfortunately, counting perfect matchings even in bipartite graphs is $\#$P-complete \cite{V79}.

The key insight of the present paper circumvents the apparent obstacle formed by the intractability of counting matchings: we only need to be able to efficiently compute \emph{some} fixed weighted sum of the matchings (with no weights set to zero). This is exactly where the determinants come to our rescue.
The price we pay is that we have to give up counting the solutions over the natural numbers.
Here we demonstrate the result through counting over fields of characteristic two which only allow us to detect if there is a cover at all and gives us little knowledge
of their number. Furthermore, to avoid having an even number of solutions cancel we will employ a 
fingerprint technique, very much in the same spirit as Williams~\cite{W09} recently extended the $k$-path detection algorithm based on an algebraic sieving
method of Koutis~\cite{K08}. The fingerprint idea is to think of the computation as evaluating a polynomial of a degree much smaller than the number of elements of its base field and then computing it at a randomly chosen point.
The fact that a polynomial cannot have more roots than its degree assure us that with great probability we discover with this single point probing whether the polynomial is the zero-polynomial or not.
We will in fact use the multivariate polynomial analogue, see e.g. \cite{MR95}.
\begin{lem}[Schwartz-Zippel]
\label{lem: sz}
Let $P(x_1,x_2,...,x_n)$ be a non-zero $n$-variate polynomial of degree $d$ over a field $F$. Pick $r_1,r_2,...,r_n\in F$ uniformly at random, then
\[
\mbox{Pr}(P(r_1,r_2,...,r_n)=0)\leq \frac{d}{|F|}
\]
\end{lem}
For now, it is sufficient to think of the inclusion--exclusion formula of Eq.~\ref{eq: inc-exc} as evaluating a multivariate polynomial over the base field GF$(2^m)$ for some $m$.
In what follows we will associate with all edges $e$ in the input hypergraph a variable $v_e$. 
Our modified version of Eq.~\ref{eq: inc-exc} reads as follows.

\begin{lem}
\label{lem: w2}
Given an X$k$C-instance $H=(V,E)$ and the family of all its solutions $\mathcal{S}$, we have that, for every subset $U\subseteq V$,
\begin{equation}
\label{eq: count}
\sum_{X\subseteq V-U} W_{2,f}(H,U,X) = \sum_{E'\in \mathcal{S}} \prod_{e\in E'} v_e^{f(e)}
\end{equation} 
where the computation is over a multivariate polynomial ring over GF$(2^m)$, $f$ is a function mapping the edges to the positive integers, and 
\begin{equation}
W_{2,f}(H,U,X)=\sum_{E''} \prod_{e\in E''} v_e^{f(e)}
\end{equation}
where the summation is over all $E''\subseteq E$, satisfying four constraints
\begin{itemize}
\item Avoidance, $\forall e\in E'':e\cap X = \emptyset$
\item Cardinality, $|E''|=|V|/k$
\item Coverage, $U\subseteq \cup_{e\in E''} e$
\item Disjointness, $\forall e_1\neq e_2\in E'': e_1\cap e_2 \cap U=\emptyset$
\end{itemize}
\end{lem}

\proof
First, note that every $E'\in \mathcal{S}$ fulfills all four conditions Avoidance, Cardinality, Coverage, and Disjointness for $X=\emptyset$, 
but violates Avoidance for every other $X$, irrespective of the choice of $U$.
Thus, the contribution $\prod_{e\in E'} v_e$ of every solution $E'$ is counted precisely once.

Second, a non-solution $E''$ obeying the three conditions Cardinality, Coverage, and Disjointness, fulfills the Avoidance condition for an \emph{even} number of choices
of $X$ irrespective of $U$, namely for all subsets of the elements of $V$ that the union of the sets in $E''$ fails to cover. Hence, all of these contributions $\prod_{e\in E''} v_e^{f(e)}$ cancel each other since we are working in a field of characteristic two.
\qed

Combining the two Lemmas above \ref{lem: sz} and \ref{lem: w2} into an algorithm choosing a random point $r_1,r_2,...,r_{|E|}\in \mbox{GF}(2^m)$ and evaluating the left-hand sum of Eq.~\ref{eq: count} in the straightforward fashion, we get:
\begin{cor}
\label{cor: alg}
Given an X$k$C-instance $H=(V,E)$ and a subset $U\subseteq V$, there is a Monte Carlo algorithm which returns ``No'' whenever there is no cover and returns ``Yes'' with probability at
least $1-\max_{e\in E} f(e)|V|/(k2^m)$ when there exists at least one, running in time $O^*(2^{|V|-|U|}\tau(W_{2,f},U))$, where $\tau(W_{2,f},U)$ is the time required to evaluate
any of the polynomials $W_{2,f}(H,U,X)$ for $X\subseteq V-U$, in a random point over the base field GF$(2^m)$.
\end{cor}

Note that by letting $m$ be in the order of $n$, when $f$ is bounded by a constant, we get exponentially low probability of failure in $n$. 
Armed with Corollary~\ref{cor: alg}, we can start looking for projections $U$ over which the computation of $W_{2,f}(H,U,X)$ is easy. The
next two sections will describe two examples of how we can use determinants to accelerate the computation.

\section{$k$-Dimensional Matching}
We begin by the easier application, $k$DM.
For this problem we can trivially find a large vertex subset on which the projected instance is a multigraph, and in fact also bipartite: we just use any two of the parts in the vertex partition given as input. 
Edmonds~\cite{E67} observed that one could relate a bipartite graphs' perfect matchings to
the determinant of a symbolic matrix. A perfect matching is a collection of disjoint edges so that every vertex is covered by precisely one edge.
To a given a bipartite graph $G=(U,V,E), n=|U|=|V|$, he associated an $n\times n$-matrix $\mathbf{A}$ with rows representing vertices in $U$, and columns the
vertices of $V$, and equated $\mathbf{A}_{i,j}$ with a variable $v_{ij}$ if $(i,j)\in E$ and zero otherwise. He showed that the determinant of $\mathbf{A}$ is non-zero
iff $G$ has a perfect matching. We will use essentially the same result, with the small exception that we need to deal with multiple edges between a vertex pair, making sure
all contributes. Formally
\begin{definition}
Given a hypergraph $H=(V,E)$ and a subset $U\subseteq V$ such that the projected hypergraph $H[U]$ is a bipartite multigraph on 
two equally sized vertex parts $U_1\cup U_2=U$, its
Edmonds matrix, denoted $\mathbf{E}(H,U_1,U_2)$, is defined by

\begin{displaymath}
\mathbf{E}(H,U_1,U_2)_{i,j}=\sum_{\substack{ e=(i,j)\in E[U] \\ i\in U_1,j\in U_2}} v_e 
\end{displaymath}
where again, $v_e$ is a variable associated with the edge $e$.
\end{definition}
We formulate our Lemma in terms of a special case of X$k$C instead of $k$DM directly to capture a more general case.
\begin{lem}
\label{lem: ed}
For a X$k$C-instance $H=(V,E)$ and two equally sized disjoint vertex subsets $U_1,U_2\subseteq V$ such that the projected hypergraph $H[U_1 \cup U_2]$ is a
bipartite multigraph,
\begin{equation}
\mbox{\emph{det}}(\mathbf{E}(H,U_1,U_2))=\sum_{M\in \mathcal{M}} \prod_{e\in M} v_e
\end{equation}
where the computation is over a multivariate polynomial ring over GF($2^m$) for some $m$ and the summation is over all 
perfect matchings $\mathcal{M}$ in $H[U_1\cup U_2]$.
\end{lem}
\proof
By definition of the determinant~\ref{eq: det}, the summation is over all products of $n$ of the matrix elements in which every row 
and column are used exactly once. Transferred to the associated bipartite graph, this corresponds to a perfect matching in the graph since rows and columns represent the
two vertex sets respectively. Moreover, the converse is also true, i.e. for every perfect matching there is a permutation describing it. Hence the mapping is one-to-one.
The inner product counts all choices of edges producing 
a matching described by a permutation $\sigma$ since:
\begin{equation}
\prod_{i=1}^n \mathbf{E}_{i,\sigma(i)} = \prod_{i=1}^n \sum_{e=(i,\sigma(i))} v_e = \sum_{M \in \mathcal{M}(\sigma)} \prod_{e\in M} v_e
\end{equation}
where $\mathcal{M}(\sigma)$ is the set of all perfect matchings $e_1,e_2,...,e_n$ such that $e_i=(i,\sigma(i))$.
\qed

\subsection{The Algorithm}
Now we are ready to prove Theorem~\ref{thm: kDM}. Given an input instance $H=(V_1,V_2,...,V_k,E)$ to the $k$DM problem where $V_1,V_2,...,V_k$ describe the vertex partition of the $n$ vertices,
we simply let $U=V_1\cup V_2$ in the algorithm
described by Corollary~\ref{cor: alg}, with $f$ mapping one to every edge. To compute $W_{2,f}(H,U,X)$ we construct the Edmonds matrix of the hypergraph $H$ \emph{restricted} to its edges
disjoint to $X$, projected on $U$, with the variables replaced by the random sample point $(r_1,r_2,...,r_{|E|})$ chosen. Next we compute its determinant.
The correctness follows from Lemma~\ref{lem: ed}, after noting that every perfect matching in a projected hypergraph contains $n/k$ disjoint edges.
The runtime bound is easily seen to be $O^*(2^{n(k-2)/k})$ since $|U|=|V_1|+|V_2|=2n/k$.

\section{Exact Cover by $k$-Sets}
\label{sec: xkc}
Next we proceed to the X$k$C problem. 
In comparison to the $k$DM we are faced with a number of additional obstacles on our way to a similar
result.
\begin{itemize}
\item First, a projection will typically capture edges differently, some will have large projections and some no at all.
\item Second, in particular the projected edges will probably not form a multigraph.
\item Third, even if they did it may not be a bipartite one.

\end{itemize}

For the first obstacle, we will prove that it is sufficient to find a projection on which at least one cover's edges all
leave projected edges of size two \emph{or less}. This is basically an extension of the idea for the X$k$C algorithm in proposition 10 in \cite{BH08}.
There, a vertex subset $U$ is picked uniformly at random of a carefully chosen size, and in the projected hypergraph only the edges which
 leave a projection of size one or less are kept. Then the inclusion--exclusion formula Eq.~\ref{eq: inc-exc} is used after noting that $W(H,U,X)$ is now
 easy to compute. The process is repeated a number of times dictated by the size of $U$. The best size to use is a trade-off of 
 the resulting summation runtime and the probability that
 a cover is projected gracefully in the sense that all its edges are kept after the projection.

For the second obstacle, in addition to handling multiple edges we also need to count perfect matchings in which loops, i.e. edges connecting a vertex to itself, count as covering the vertex of its endpoints. Since this means that not all perfect matchings will involve the same number of edges, we have to take special care to make the determinants useful.  We use polynomial interpolation to solve for the contributions of matchings of the same size separately to be able to fulfill the Cardinality constraint for $W_{2,f}$ in Corollary~\ref{cor: alg}. 
To this end we introduce an auxiliary variable $s$ parametrizing the matrices and use several determinant calculations.

For the third obstacle, we will use a variation of a result generalizing Edmonds' due to Tutte~\cite{T47}. He showed 
that even for general not necessarily bipartite graphs one can make a connection between its perfect matchings and the 
determinant of a symbolic matrix, although twice as large matrices in both directions are required. To a given a graph $G=(V,E), n=|V|$ he associates an $n\times n$-matrix
$\mathbf{A}$ with rows and columns representing the vertices, and assigns $\mathbf{A}_{i,j} = v_{i,j}$ for $i<j$ and $\mathbf{A}_{i,j} = -v_{i,j}$ for $i>j$
with $v_{i,j}$ a variable for each edge $(i,j)\in E$. The remaining entries are set to zero. The determinant of $\mathbf{A}$ is non-zero iff $G$ has a perfect matching.

We define matrices similar to Tutte's:

\begin{definition}
\label{def: tu}

Given a hypergraph $H=(V,E)$ and a subset $U\subseteq V$ such that in the projected hypergraph $H[U]$ all edges have size at most two, its
Tutte matrix of index $s$, denoted $\mathbf{T}^{(s)}(H,U)$, is defined by
\begin{displaymath}
\mathbf{T}^{(s)}(H,U)_{i,j}=\left\{\begin{array}{ll} \sum v_e & :e\in E[U], e=(i,j), i\neq j\\ 
s\sum v_e & :e\in E[U], e=(i,j), i=j\end{array}\right.
\end{displaymath}
\end{definition}

\begin{lem}
\label{lem: tu}
For a X$k$C-instance $H=(V,E)$ and a vertex subset $U\subseteq V$ such that in the projected hypergraph $H[U]$, every
edge has size at most two,
\begin{equation}
\mbox{\emph{det}}(\mathbf{T}^{(s)}(H,U))=\sum_{M\in \mathcal{M}} s^{\Lambda(M)} \prod_{e\in M} v_e^{p(e)}
\end{equation}
where the computation is over a multivariate polynomial ring over GF($2^m$) for some $m$, the summation is over all 
perfect matchings $\mathcal{M}$ in $H[U]$, $\Lambda(M)$ is the number of loops in the matching $M$, and $p(e)=1$ if $e$ is a loop and $p(e)=2$ otherwise.
\end{lem}
\proof
By definition of the determinant~\ref{eq: det}, the summation is over all products 
$
\prod_{i=1}^n \mathbf{T}^{(s)}_{i,\sigma(i)}
$
for a permutation $\sigma$. Call a permutation $\sigma$ good if $\forall i: \sigma(\sigma(i))=i$ holds, and bad otherwise. We will argue that only good permutations contribute to the sum.
To see why, consider a bad $\sigma$. Then there exists a smallest $i$ such that $\sigma(\sigma(i))=j\neq i$.
Look at the cyclic sequence $\{c_i\}$ where $c_0=i$ and $c_{k+1}=\sigma(c_{k})$ for $k>0$. Let $L>2$ be the smallest positive integer such that
$c_L=i$ (Note that there must be one and that all $c_i$ in between must by distinct since every element in $1$ through $n$ is mapped to exactly once).
Next define a cycle reversal operation $D$ mapping bad permutations on bad permutations by 
letting $D(\sigma)$ be identical to $\sigma$ except in the points $c_1$ through $c_{L}$, where instead $D(\sigma)(c_i)=c_{i-1}$.
Now first observe that the reversal operation is dual in the sense that $D(D(\sigma))=\sigma$ and that $D(\sigma)\neq \sigma$ since $L>2$, and hence every bad permutation can be uniquely paired with another bad permutation.
Second note that the contribution of a bad permutation is identical to the contribution of its dual, since the Tutte matrices are symmetrical.
Thus, since we are counting in a field of characteristic two, they cancel each other.

Next we continue to observe that the good permutations describe precisely the structure of all possible perfect matchings in a multigraph: $i$'s such that
$\sigma(i)\neq i$ describe ordinary two-vertex edges in the matching, and $i$'s such that $\sigma(i)=i$ describe loops.

The inner product of Eq.~\ref{eq: det} reads
\begin{equation}
\prod_{i=1}^n \mathbf{T}^{(s)}_{i,\sigma(i)} = \left( \prod_{i,i=\sigma(i)} s\sum_{e=(i,i)} v_e \right)\left( \prod_{i,i\neq \sigma(i)} \sum_{e=(i,\sigma(i))} v_e\right) =  \sum_{M \in \mathcal{M}(\sigma)} s^{\Lambda(M)}\prod_{e\in M} v_e
\end{equation}
where $\mathcal{M}(\sigma)$ is the set of all \emph{directed} perfect matchings $e_1,e_2,...,e_n$ described by the good permutation $\sigma$ for which $e_i=(i,\sigma(i))$.

Now consider a directed perfect matching $e_1,e_2,...,e_n$ such that for some $j$, $e_j\neq e_{\sigma(j)}$, and refer to it as being bad. We will see that all of these cancel in very much the same way as the bad permutations did.
Namely, again find the smallest $j$ for which this is the case, and define a reversal operation $R_\sigma$ mapping bad directed perfect matchings onto themselves by
exchanging $e_j$ and $e_{\sigma(j)}$. Since this operation pairs up the bad directed perfect matchings ($R_\sigma(\{e_i\})\neq \{e_i\}$ and $R_\sigma(R_\sigma(\{e_i\}))=\{e_i\}$) and we work in a field of characteristic two, their contributions cancel.
Thus we are left with only good permutations and good directed perfect matchings. The latter can be thought of as undirected perfect matchings in which every non-loop edge
is included twice in the product.
\qed

To find the contributions of matchings of the same size separately, think of the matchings partitioned in groups $\mathcal{M}_0, \mathcal{M}_1,...,\mathcal{M}_n$ according to the number of loops of the matching.
We can rewrite the determinant in Lemma \ref{lem: tu} as
\begin{equation}
\label{eq: s}
\mbox{det}(\mathbf{T}^{(s)}(H,U))=\sum_{i=0}^{n} s^{i} M_i
\end{equation}
where $M_i=\sum_{M\in \mathcal{M}_{i}} \prod_{e\in M} v_e^{p(e)}$ are the quantities we seek. The right hand side of Eq.~\ref{eq: s} is a 
degree $n$ polynomial in $s$ and thus we can solve for $M_0,M_1,...,M_n$ by computing $\mbox{det}(\mathbf{T}^{(s)}(H,U))$ in $n$ different choices of $s$,
and use Lagrange's interpolation formula to recover the sought values. In fact, either there are no matchings with an odd number of loops or no matchings with an even number of loops depending on the parity of $|U|$. Consequently, the evaluation of $n/2$ points suffices, but we disregard from this optimization possibility for simplicity.
Once we have the $M_i$'s we are close to be able to compute $W_{2,f}$ efficiently according to the following Lemma:
\begin{lem}
\label{lem: conv}
Given a X$k$C instance $H=(V,E)$ and a $U\subseteq V$ such that for all edges $e\in E$,$|e\cap U|\leq 2$, $f(e)=2$ if $|e\cap U|=2$ and $1$ otherwise, and any $X\subseteq V-U$,
\begin{equation}
\label{eq: conv}
W_{2,f}(H,U,X)=\sum_{i=0}^{|U|} Z(\frac{|V|}{k}-\lfloor \frac{|U|+i}{2} \rfloor) M_i 
\end{equation}

where $M_i=\sum_{M\in \mathcal{M}_{i}} \prod_{e\in M} v_e^{p(e)}$ are the contribution of all matchings $\mathcal{M}_i$ containing exactly $i$ loops in the projected hypergraph of $H$ on $U$ restricted 
to the edges disjoint to $X$, and
\begin{equation}
Z(i)=\sum_{\substack{E'\subseteq Z\\|E'|=i }} \prod_{e'\in E'} v_{e'}
\end{equation}
with $Z$ defining the set of edges $e$ disjoint to $X$ also having an empty intersection with $U$.
\end{lem}

\proof
The $M_i$'s count the contribution of all ways to cover $U$ with the edges which leaves a non-empty projection on $U$ and the $Z(i)$'s count the contribution of all ways to choose
edges leaving an empty projection. Note that a matching from $\mathcal{M}_i$ involves exactly $\lfloor \frac{|U|+i}{2} \rfloor$ edges if it exists. The right hand side of 
Eq.~\ref{eq: conv} convolutes over all ways their total number of edges could equal $|V|/k$ in order to meet the Cardinality constraint in Corollary~\ref{cor: alg}.
\qed

The only piece missing is a simple way to evaluate $Z(i)$, and we note that it can be done by dynamic programming through a simple recursion.
Number the edges in $Z$ defined in Lemma~\ref{lem: conv} arbitrarily as $e_1,e_2,...,e_p$, set $Z_i=\{e_1,e_2,...,e_i\}$, and define
\begin{equation}
z(i,j)=\sum_{\substack{E'\subseteq Z_j\\ |E'|=i}} \prod_{e'\in E'} v_{e'}
\end{equation}
These can be solved for by
\begin{equation}
\label{eq: dp}
z(i,j)=\left\{ \begin{array}{cl} 1 & : i=j=0 \\ 0 & : i=0 \mbox{ or } j=0 \\ z(i-1,j-1)v_{e_j}+z(i,j-1) & : \mbox{otherwise}\\ \end{array} \right.
\end{equation}
and we finally compute $Z(i)$ through $Z(i)=z(i,p)$.

\subsection{The Algorithm}
\label{sec: xkcalg}
We are ready to prove Theorem~\ref{thm: XkC}. First we describe the algorithm.
Given an input instance $H=(V,E)$ to the X$k$C problem, we compute two parameters $t$ and $I$ depending on $k$. These are given by the calculations
in the next section~\ref{sec: analysis}. We repeat the following procedure until we detect a cover, in which case we report so, or have tried unsuccessfully $I$ times, in which case we report that no cover was found:

\begin{algo}
\label{algo: proj on U}
\hspace{5mm}
\begin{enumerate}
\item Choose a $tn$-sized subset $U\subseteq V$ uniformly at random.
\item Construct $H_U=(V,E_U)$ where $E_U=\{e|e\in E,|e\cap U|\leq 2\}$.
\item Run the summation algorithm in Corollary~\ref{cor: alg} on $H_U$, using $U$, and  let $f(e)=2$ if $|e\cap U|=2$ and $1$ otherwise. 
Use the method of the previous section~\ref{sec: xkc} to compute $W_{2,f}(H_U,U,X)$, i.e.
\begin{enumerate}
\item Construct the Tutte matrices $\mathbf{T}^{(s)}$ of $H_U[U]$ restricted to its edges which are disjoint to $X$ for $s=g^i,0\leq i \leq |U|$ where $g$ is a generator of the multiplicative group in GF($2^m$). 
\item Compute the determinants of $\mathbf{T}^{(s)}$.
\item Use Lagrange interpolation to solve for the $M_i$'s via Eq.~\ref{eq: s}.
\item Calculate the $Z(i)$'s by Eq.~\ref{eq: dp}.
\item Evaluate Eq.~\ref{eq: conv}.
\end{enumerate}
\end{enumerate}
\end{algo}

Given that the random $U$ is such that all edges in \emph{some} exact cover $S$ are kept in $H_U$, the previous Section~\ref{sec: xkc} verifies its correctness:
Lemmas~\ref{lem: tu} and \ref{lem: conv} together with the observation that $g^i$ for $1\leq i \leq |U|$ are all distinct points, assures us that step (3) of the algorithm works. 
We are left with deciding $t$ and $I$ to make it very likely that some exact solution is kept at least once and tune them to get the best possible runtime.

\subsection{Runtime Analysis}
\label{sec: analysis}
Our runtime analysis hinges on the probability that any fixed solution $S$ to the X$k$C instance $H=(V,E)$ when projected on a subset $U\subseteq V$
chosen uniformly at random from the $tn$-sized subsets of $V$ for some fraction $t$ of the vertices, gets all its edges to leave a small projection on $U$, namely $\forall e\in S:|e\cap U|\leq 2$.
We denote this event by $\varepsilon(t)$.
If we repeat the process $I$ times, the probability that none of the $I$ independent
random selections for $U$ is successful in the sense that they retain $S$ after the projection,
is at most $(1-\mbox{Pr}(\varepsilon (t)))^I < e^{-\mbox{Pr}(\varepsilon (t))I}$. Consequently, we need $I=\log(\epsilon^{-1}) \mbox{Pr}(\varepsilon(t))^{-1}$ 
to get probability at least $1 - \epsilon$ for one or more of the $I$ selections to be successful.
Thus we may use $\epsilon = c^{-n}$ for some constant $c >1$ to get an exponentially low probability in $n$ of failure without increasing the number of
repetitions $I$ by more than a polynomial factor.

To bound the probability of the event, we count the number of good $tn$-sized subsets of the vertices. This is a binomial sum (actually a trinomial one) over the number of
edges in the solution $S$ which gets a projection of size two:
\begin{equation}
\sum_{t_1+2t_2=tn} \binom{n/k}{t_1}\binom{n/k-t_1}{t_2}\binom{k}{2}^{t_2}\binom{k}{1}^{t_1}
\end{equation}

To lower bound this sum of all non-negative terms, we will use just one of them. Let $N=n/k$ and parametrize $tn=\tau_{12}N+\tau_2N$ where $\tau_{12}$ is the fraction of
sets in the solution $S$ which gets at least one of its elements chosen, and $\tau_2$ is the fraction of sets that gets two. 

Then, we bound our probability as the quotient of the single term lower bound on the number of good sets and the number of all sets $\binom{n}{tn}$ to
\begin{equation}
\label{eq: pr}
\mbox{Pr}(\varepsilon(t))\geq \frac{\binom{N}{\tau_{12}N}\binom{\tau_{12}N}{\tau_2N}k^{\tau_{12}N}(k-1)^{\tau_2N}}{2^{\tau_2N}\binom{kN}{(\tau_{12}+\tau_2)N}}
\end{equation}

The runtime of Corollary~\ref{cor: alg} is $O^*(2^{n-tn})$ given our polynomial time algorithm for computing $W_{2,f}$. Omitting polynomial factors, Algorithm~\ref{algo: proj on U} for X$k$C has to run for $Pr(\varepsilon(t))^{-1}$ different choices of $U$ in the worst case. Let $T_{k,t}$ denote the final runtime, and expand the binomials of Eq.~\ref{eq: pr} to get:
\begin{equation}
T_{k,t} \leq \frac{2^{n-tn}}{\mbox{Pr}(\varepsilon(t))} \leq \frac{2^{kN-\tau_{12}N}(\tau_2N)!(N-\tau_{12}N)!(\tau_{12}N-\tau_2N)!(kN)!}{N!k^{\tau_{12}N}(k-1)^{\tau_2N} (kN-\tau_{12}N-\tau_2N)!(\tau_{12}N+\tau_2N)!}
\end{equation}

\begin{center}
    \begin{table}
    \begin{tabular}{ | l | l | l | l | l | l |}
    \hline
   $k$  & $\tau_{12}$ &  $\tau_2$ &   $t$     &$I^{1/n}$ & $c_k$  \\ \hline
     3  &   0.961  &  0.679 &  0.547    & 1.092    & 1.496  \\ \hline
     4  &   0.936  &  0.613 &  0.387    & 1.073    & 1.642  \\ \hline
     5  &   0.921  &  0.583 &  0.301    & 1.060    & 1.721  \\ \hline
     6  &   0.912  &  0.565 &  0.246    & 1.050    & 1.771  \\ \hline
     7  &   0.905  &  0.554 &  0.208    & 1.043    & 1.806  \\ \hline
     8  &   0.900  &  0.546 &  0.181    & 1.038    & 1.832  \\ \hline
    \end{tabular}
      
    \caption{ Numerically found parameters $\tau_{12}$ and $\tau_2$ which approximately minimizes $c_k$.}
    \label{tab: params}
    \end{table}
\end{center}

If we replace the factorials with Stirling's approximation $n!\in \theta(\sqrt{n}(n/e)^n)$ and divide $(N/e)^{k+1}$ out of both numerator and denominator, 
we are left with a slightly less intimidating expression

\begin{equation}
\label{eq: rtbound}
T_{k,t} \leq \left(\frac{2^{(k-\tau_{12})}\tau_2^{\tau_2}(\tau_{12}-\tau_2)^{\tau_{12}-\tau_2}(1-\tau_{12})^{1-\tau_{12}} }{k^{\tau_{12}-k}(k-1)^{\tau_2}(k-\tau_{12}-\tau_2)^{k-\tau_{12}-\tau_2}(\tau_{12}+\tau_2)^{\tau_{12}+\tau_2}}\right)^N
\end{equation}

Rewriting this as $T_{k,t}\leq c_k^n$ we see that $c_k$ can be obtained as the $k$:th root of the expression within the brackets in Eq.~\ref{eq: rtbound}.
Solving numerically for the choices of $\tau_{12}$ and $\tau_2$ that minimizes $c_k$ we find that the minimum moves slightly with increasing $k$, see Table~\ref{tab: params}.
The minimum, however, lies in a quite flat neighborhood within a large vicinity of the actual minimum, and comparable bounds not too far from the best possible with our technique are obtained with fixed parameters for all $k$ by, say, $\tau_{12}=0.9$ and $\tau_2=0.6$. With this choice
of parameters in Eq.~\ref{eq: rtbound} we obtain the general bound in Theorem~\ref{thm: XkC}.

\section*{Acknowledgements}
This research was supported in part by the Swedish Research Council project "Exact Algorithms".

\end{document}